**From virtual patients to digital twins in immuno-oncology: lessons learned from mechanistic quantitative systems pharmacology modeling**

Hanwen Wang[1*], Theinmozhi Arulraj[1], Alberto Ippolito[1], and Aleksander S. Popel[1,2]

[1] Department of Biomedical Engineering, Johns Hopkins University School of Medicine, Baltimore, MD 21205, USA

[2] Departments of Medicine and Oncology, and the Sidney Kimmel Comprehensive Cancer Center, Johns Hopkins University School of Medicine, Baltimore, MD 21205, USA

* Corresponding author
E-mail: hwang163@jhmi.edu

**Abstract**

Virtual patients and digital patients/twins are two similar concepts gaining increasing attention in health care with goals to accelerate drug development and improve patients' survival, but with their own limitations. Although methods have been proposed to generate virtual patient populations using mechanistic models, there are limited number of applications in immuno-oncology research. Furthermore, due to the stricter requirements of digital twins, they are often generated in a study-specific manner with models customized to particular clinical settings (e.g., treatment, cancer, and data types). Here, we discuss the challenges for virtual patient generation in immuno-oncology with our most recent experiences, initiatives to develop digital twins, and how research on these two concepts can inform each other.

**Introduction**

Cancer is a leading cause of death worldwide with the lowest success rate of clinical trials among all complex diseases. In an analysis of clinical trials from 2000 to 2015, the overall probability (defined by Wong et al.[1]) of a drug successfully moving from phase I to approval was merely 3.4% in oncology, but for trials that used biomarkers for patient selection, the probability of success was significantly improved[1]. With increasing number of newly discovered drugs and potential biomarkers to investigate, it is extremely difficult to test and compare all dose levels, therapy combinations, and predictive biomarkers for each cancer type via clinical trials, which necessitates development of computational tools to accelerate the process.

Since 1990s, mathematical models have and continue to play important roles in drug development as a cost-efficient tool to inform clinical trial design (i.e., model-informed drug development or MIDD)[2]. Semi-mechanistic approaches like pharmacokinetic-pharmacodynamic (PKPD) models started first to accompany regulatory submissions, and, with advancing mechanistic understanding of pathophysiology and increasing computational power, mechanistic models, such as physiologically-based pharmacokinetic (PBPK) and quantitative systems pharmacology (QSP) models, were developed[2,3]. From 2013 to 2020, the US Food and Drug Administration has received a rising number of new drug applications with the support of QSP models, more than one fifth of which were for oncologic diseases[4]. Therefore, hypothesis-driven mechanistic QSP modeling has begun to play a critical role in predicting effectiveness of newly discovered drugs and determining the optimal dosage to assist clinical trial design via clinical trial simulation (i.e., in silico/virtual clinical trial)[4–7] with virtual patients.

Virtual patients are usually defined as model parameterizations that generate physiologically plausible outputs[8]. Parameters with physiological or biological definitions should also be confined by the experimentally and clinically observed values. By generating a virtual patient population with similar characteristics to the target patient cohort, mechanistic models can compare different therapy combinations and potential biomarkers for patient stratification[9]. In the past few years, virtual patients in immuno-oncology are commonly generated via random sampling from chosen distributions[10,11] or by models whose variables can be relatively easy to measure in clinical settings, such as imaging-based models[12]. Nonetheless, the strong inter-patient, inter-tumoral, and intra-tumoral heterogeneities in cancer require large clinical datasets to determine the physiological plausibility of randomly generated virtual patients. This challenge may be resolved by emerging multi-omics data[13,14], which involve a large number of molecular data that characterize the tumor microenvironment in individual patients. In parallel with the effort to generate virtual patients that resemble real patients' characteristics, digital twins are being developed in precision oncology with a goal to monitor and optimize treatment for individual patients through personalized models[15]. In this perspective, we aim to share our experience in generating virtual patients based on multi-omics analyses and discuss current efforts in developing realistic virtual patients and digital twins, which share a similar definition but are usually generated for different goals in immuno-oncology.

**Virtual patients for clinical trial simulation**

As each virtual patient is represented by a unique set of model parameterization[16], it is critical to define the physiologically plausible ranges for model parameters and variables. In immuno-oncology, this is particularly challenging because: 1) the mechanisms involved in the cancer-immunity cycle[17,18], including mechanism of action (MoA) of newly discovered therapeutics, are sometimes not fully understood (thus their effects are approximated by Hill functions rather than detailed biochemical equations), 2) variables, such as immune cell densities and cytokine levels, are subject to strong inter-individual and spatio-temporal heterogeneity, and 3) most of these parameters and variables are either unavailable, or difficult to measure, allometrically scaled from animal models, or only measured at 1-3 time points throughout a clinical study. Perhaps the most studied therapeutic area in virtual patient generation is cardiovascular diseases, where measurements (e.g., heart rate, cardiac output) are relatively easy to take and mechanisms are better understood than those in cancer[19]. Despite the challenges in immuno-oncology, Craig and colleagues have summarized the steps to conduct virtual clinical trials[11] and demonstrated the approaches to generate virtual patients using a tumor growth model[10]. Cheng et al. also discussed key considerations during development of QSP models and virtual patient algorithms[16]. In parallel, our group has also applied various virtual patient generation methods to a quantitative systems pharmacology model for immuno-oncology (QSP-IO)[20], in subsequent models, using data from multiplex digital pathology and genomic analysis. In this section, we discuss the challenges met during the QSP-IO model development.

A sequence of QSP-IO models with progressively more detail of the tumor immune microenvironment (TiME), such as cell types and cytokines, have been developed by our group with a goal to predict effectiveness of immune checkpoint inhibitors in combination with other therapies in multiple cancer types[20–27]. The latest model version was used for a retrospective analysis of anti-PD-L1 treatment in non-small cell lung cancer[28], as well as a prospective prediction for the effectiveness of a masked antibody in triple-negative breast cancer[29]. In another study of metastatic triple-negative breast cancer, the model was expanded to account for up to 3 metastatic lesions, which were parameterized to transcriptomic data from lung and other metastases (i.e., liver, and bone) of breast tumors[30]. Furthermore, to account for spatio-temporal heterogeneity, the QSP model was also integrated with an agent-based model (spQSP-IO), which was calibrated by multiplex digital pathology[31–33] and spatial transcriptomics[34].

As the first step of generating a virtual patient population, a subset of model parameters is selected that best represent the inter-individual heterogeneity and randomly generate their values via Latin Hypercube Sampling. Some studies assume uniform distribution for all parameters (i.e., no prior information) and set an upper and a lower boundary to randomly generate parameter values, relying on subsequent algorithms to filter out virtual patients whose model-predicted characteristics (e.g., blood volume, heart rate in a cardiovascular disease model; tumor size, T cell level in a tumor growth model) fall out of the physiologically plausible range[35,36]. In QSP-IO, however, the main outputs of interest are: 1) tumor size for response status prediction, and 2) immune profiles, such as intratumoral CD8, CD4, FoxP3 T cell density, and receptor/ligand expression level, such as PD-L1 and CTLA-4, for biomarker analysis[28]. These variables often have wide ranges observed in patients[37–39]. For example, in a digital pathology analysis of tumor tissue samples from 43 patients with breast cancer, the density of CD8 T cells can differ with more than 3 orders of magnitude[38]. Therefore, filtering out virtual patients based on their T cell levels is not an effective method especially when other clinical measurements (e.g., cytokine profile) are unavailable to further narrow down the physiologically plausible domain.

In our model and other QSP studies in immuno-oncology, parameter distributions are often estimated by published experimental or clinical data[27–30,40,41]. Lognormal distribution is commonly assumed for physiological/biological parameters[42]. Parameters that cannot be directly measured or have limited availability from the literature were calibrated by iterations of clinical trial simulation. For each iteration, at least 1,000 virtual patients were randomly generated to calculate the outputs of interest. Then, parameters were adjusted by comparing medians of model outputs to clinically measured values. This is a time-consuming step but necessary due to the nonlinear nature of the model, in which case, median parameter values do not correspond to median model output values.

QSP models usually consist of hundreds of cellular/molecular species. Therefore, it is challenging to find the initial conditions for all the model variables, which correspond to the patient status at the beginning of the drug administration. To this end, with each randomly sampled parameter set, we first initialize the model with a single cancer cell, a baseline level of cytokines, naïve T cells, antigen-presenting cells, and cell surface molecules (i.e., immune checkpoints, major histocompatibility complex (MHC), co-stimulatory ligands and receptors), and set the other variables to zero. Measurements from healthy individuals can assist estimation of these baseline patient characteristics[43,44]. In addition, a pre-treatment tumor size is randomly assigned to each virtual patient. At the end of the simulation, the model outputs at the time point when the pre-treatment tumor size is reached are considered the patient's pre-treatment characteristics, which set the initial conditions for the clinical trial simulation.

With the unprecedented increase in omics data on specific cancer types from collaborative studies, such as the TCGA[45], AURORA[46], Human Tumor Atlas Network (HTAN)[47], and iAtlas[48], it is possible to use the immune cell proportions derived from omics data for virtual patient generation. In our recent QSP studies[28,29], we selected virtual patients whose pre-treatment characteristics statistically matched the patient data using the Probability of Inclusion by Allen et al[35]. Figure 1A compares the distribution of three pre-treatment immune subset ratios in the plausible and virtual patients from our NSCLC model[28] with that in the real patient data on lung adenocarcinoma (LUAD) and squamous cell carcinoma (LUSC) from iAtlas database. Plausible patients (i.e., virtual patient candidates) refer to those randomly generated by calibrated parameter distributions, from which a virtual cohort was selected by the Probability of Inclusion. In short, the Probability of Inclusion ($P_i$) of a plausible patient that corresponds to a model output $\hat{y}$ was proportional to the ratio between the multivariate probability density function of the real patient data ($p_D$) and that of the plausible patient cohort ($p_{pl}$) at $\hat{y}$ (i.e., $P_i \propto p_D(\hat{y})/p_{pl}(\hat{y})$)[35]. When compared separately, the distributions of each immune subset ratio in the virtual patient cohort are statistically the same as those in the real patient data according to Kolmogorov-Smirnov test (Figure 1A).

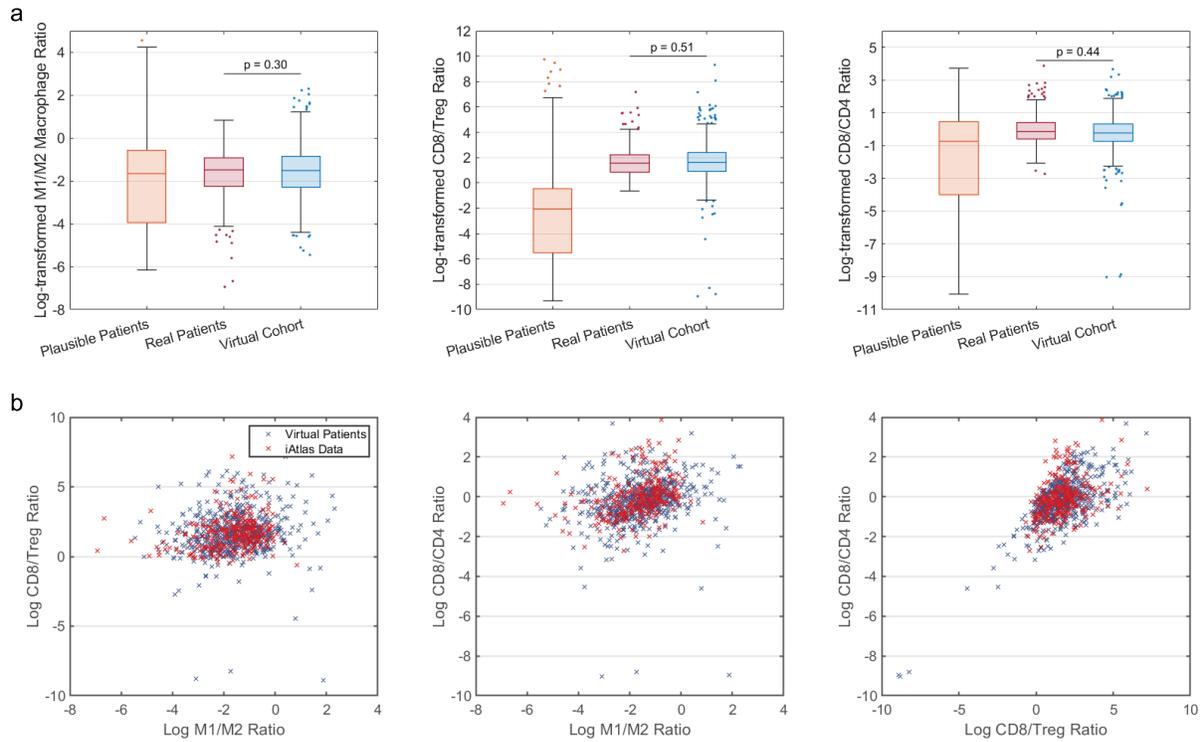

**Figure 1. Comparison of immune cell subset ratio distributions and correlations between the virtual patients and real patient data.**

Immune cell subset ratios were calculated by immune cell proportions from the iAtlas database. Natural-log transformation was performed. **a** The 25th, 50th, and 75th percentiles are encoded by box plots with whiskers that define 1.5 times the interquartile range away from the bottom or top of the box. **b** Correlations between each pair of immune subset ratios in the virtual population and the real patient data.

Figure 1B shows the correlations between each pair of the three ratios (i.e., CD8/CD4, CD8/Treg, and M1/M2 tumor-associated macrophages) in the virtual patients and the iAtlas data. When compared using Fisher's transformation and Z-test, the correlation coefficient between CD8/CD4 and CD8/Treg ratios in the virtual patients is significantly different from that in the iAtlas data, while the correlations between the other two pairs are similar. We hypothesized that this difference is due to the outliers in the plausible patient cohort, particularly those with low CD8/Treg and CD8/CD4 ratios in Figure 1B, which can be over-selected by the Probability of Inclusion due to their low probability density in the randomly sampled plausible patient cohort $(p_s \rightarrow 0)$ yet finite $p_d$.

Omics data also allowed us to investigate the behavior of metastatic lesions. Given the limited availability of immune cell abundance data for metastatic tumors directly from the literature or existing databases, we used RNA-seq data on breast cancer metastases generated by Siegel et al.[49] in our recent study of metastatic breast cancer[30]. Using immune cell deconvolution methods EPIC[50] and quanTIseq[51], we were able to estimate the immune cell proportions and calculate the ratios of immune cell subset proportions between primary breast tumors and specific metastatic sites (e.g., lung, liver, and bone). We then used the abundance estimates of T cells, dendritic cells, and macrophages from lung, liver, and bone metastases of breast tumors to inform virtual patient generation. More specifically, distributions of virtual patient parameters, such as the recruitment rates of immune cells to metastatic tumors, were calibrated such that the ratios of immune cell subset level between the metastatic lesions and the primary tumor in the model were consistent with estimates obtained from RNA-seq data using deconvolution methods.

**Interplay between virtual patients and digital twins**

Another concept related to virtual patients is digital twins, whose definition may differ across research fields. In drug development, virtual patients aim to predict effectiveness of treatments on the population level, such as newly discovered drugs and novel drug combinations, and inform future clinical trial design[16]. On the other hand, health or medical digital twins are often developed for personalized medicine, which emphasize the correspondence with their physical counterparts in the real world[52–54]. In practice, virtual patients can be generated from data compiled from multiple patients (e.g., summary statistics), which provides a realistic representation of a complex disease and allows clinical trial simulation to make population level predictions and assist drug developers[55]. Digital patients/twins are often generated by measurements from individual patients, assuming periodical updates with new measurements in the future and aiming to find the best option of patient care for clinicians[56]. Therefore, while digital twin is an individualized patient model, virtual patient could be a more generic patient model.

Wright and Davidson[57] discussed optimal properties of a model and data required to build digital twins. The model should: 1) include sufficient mechanistic details so that parameters can be periodically updated with new measurement data; 2) make accurate predictions with new parameter values; and 3) be computationally efficient to make on-time predictions for its purpose. The authors also discussed the possibility of reducing the model to describe a local part of a complete system, or replacing a subpart of the model with a surrogate model (e.g., a data-driven model) to, for example, increase its computational efficiency[57]. When adopting these principles to build medical digital twins, An and Cockrell[58] pointed out the challenges, including the lack of mechanistic understanding in biology, difficulties to acquire quantitative data to periodically update model parameters, and measurement uncertainties commonly observed in biological experiments. Meeting these challenges, Laubenbacher et al. proposed a 4-stage method specifically for generating digital twins of the human immune system[59].

In the past few years, digital twins have gained attention in immuno-oncology research in parallel with virtual patient development[15,60–62]. Given the stricter requirements for digital twins, models are usually designed to accommodate the type of available data. For example, Jarrett et al. designed a 3-D mathematical model of tumor growth and response to treatments, which was governed by a partial differential equation, allowing for model calibration by routine MRI data from individual patients[63]. It was applied to generate digital twins for patients with triple-negative breast cancer to predict patient-specific response to a neoadjuvant chemotherapy[60]. The model-predicted post-treatment tumor size change from baseline and rate of pathological complete response were compared with the ARTEMIS trial for model validation by concordance correlation coefficients and the area under the receiver operator characteristic (auROC), respectively (pre-treatment and on-treatment MRI data used for digital twin calibration). The authors also observed different uncertainty levels in model predictions, and suggested that more frequent on-treatment measurements are warranted to periodically update digital twins, which is an important digital twin technique to increase model accuracy and lower uncertainty in model predictions[60]. Notably, the feedback loop between each patient's new measurements and their digital twin model is a meticulous process and requires careful selection of methods for parameter inference, uncertainty quantification, etc.[64]

Although recent research on digital twins and virtual patients are mostly independent of each other potentially due to different study goals, it is important to note that they share similar challenges, and therefore one approach that tackles the challenges can be applicable to both fields[59]. Both virtual patients and digital twins are limited by the lack of sufficient patient data. Although summary statistics can provide guidance on generating physiologically plausible virtual patients, patient-specific data retain the information on correlations between variables (as demonstrated above), thus improving the model's predictive power[28,29]. When time-course data (e.g., tumor size measurements over time) become available, virtual patients generated by pre-treatment data can be validated by subsequent measurements. Time-course data would also allow digital twin models to evolve with the patients and improve mechanistic understanding when comparing model prediction with clinical observation iteratively[59,65]. In addition, by treating a virtual patient population as digital twin candidates, digital twins can be derived for new patients by selecting the virtual patients whose model-predicted profiles match the time-course data (i.e., match virtual patients with their physical counterparts in the real world to become digital twins). In this way, model parameters of a digital twin no longer need to be directly derived from clinical measurements, and multiple digital twin candidates can be selected from a virtual patient population to quantify uncertainties in model predictions, such as in the analysis of non-Hodgkin lymphoma by Susilo et al[66].

When digital twin models are built on data collected from a specific patient group, they can be later repurposed to generate a virtual patient population that shares similar traits, which can be used to make population level predictions, such as the effectiveness of newly discovered drugs targeting the same patient group. Tivay et al. suggested using compressed latent parameterization to generate realistic virtual patients from parameter values fitted to real patient data (i.e., digital twins)[67]. After constructing a latent parameter space from virtual patients fitted to patient-specific data (i.e., digital twins), new virtual patients can be generated from the latent space. In fact, we have applied this approach to account for the inter-patient heterogeneity in pharmacokinetic (PK) parameters in our NSCLC study[28], which resulted in PK profiles in the virtual patient population that were comparable to that in real patients. Given the sparse public data that are available from oncology trials, methodology research on generating new realistic virtual patients from digital twins can provide an alternative approach to challenge.

**Conclusion**

Generating physiologically plausible virtual patients is critical for making accurate prediction of the effectiveness of novel treatments using mechanistic models. We have discussed the current challenges faced by modelers and our approaches to these challenges. We emphasize the importance of public databases like iAtlas that contain correlated variables on the patient level (e.g., data corresponding to each patient). Further, since new methods created to generate virtual patients and medical digital twins are usually demonstrated in cardiovascular diseases and fields other than immuno-oncology[68], this field can benefit from future research on the applicability and robustness of methods that have been proven useful in other therapeutic areas. Moreover, we raise the possibility of generating digital twins from virtual patients and vice versa, facilitating collaboration between researchers in these two fields.

## Acknowledgements

This work is supported in part by NIH grant R01CA138264.

## Competing Interests

ASP is a consultant to Incyte, J&J/Janssen, and AsclepiX Therapeutics. The terms of these arrangements are being managed by the Johns Hopkins University in accordance with its conflict-of-interest policies. A.I. is a current employee of AstraZeneca PLC. The remaining authors declare no competing financial or non-financial interests.

## Author Contributions:

All authors conceptualized this study. A.S.P. supervised this study and acquired funding. H.W., T.A., and A.I. designed the methodology. H.W. analyzed the results and wrote the manuscript. All authors reviewed and revised the manuscript.

## References

1. Wong, C. H., Siah, K. W. & Lo, A. W. Estimation of clinical trial success rates and related parameters. *Biostatistics* **20**, 273–286 (2019).

2. Madabushi, R., Seo, P., Zhao, L., Tegenge, M. & Zhu, H. Review: Role of Model-Informed Drug Development Approaches in the Lifecycle of Drug Development and Regulatory Decision-Making. *Pharm Res* **39**, 1669–1680 (2022).

3. Azer, K. *et al.* History and Future Perspectives on the Discipline of Quantitative Systems Pharmacology Modeling and Its Applications. *Front. Physiol.* **12**, 637999 (2021).

4. Bai, J. P. F. *et al.* Quantitative systems pharmacology: Landscape analysis of regulatory submissions to the US Food and Drug Administration. *CPT Pharmacom & Syst Pharma* **10**, 1479–1484 (2021).

5. Holford, N. H. G., Kimko, H. C., Monteleone, J. P. R. & Peck, C. C. Simulation of Clinical Trials. *Annu. Rev. Pharmacol. Toxicol.* **40**, 209–234 (2000).

6. Brown, L. V., Gaffney, E. A., Wagg, J. & Coles, M. C. Applications of mechanistic modelling to clinical and experimental immunology: an emerging technology to accelerate immunotherapeutic discovery and development. *Clinical and Experimental Immunology* **193**, 284–292 (2018).

7. Sorger, P. K. *et al.* Quantitative and systems pharmacology in the post-genomic era: new approaches to discovering drugs and understanding therapeutic mechanisms. *An NIH White Paper by the QSP Workshop Group* (2011).

8. Michelson, S. The impact of systems biology and biosimulation on drug discovery and development. *Mol. BioSyst.* **2**, 288 (2006).

9. Chelliah, V. *et al.* Quantitative Systems Pharmacology Approaches for Immuno-Oncology: Adding Virtual Patients to the Development Paradigm. *Clin Pharma and Therapeutics* **109**, 605–618 (2021).

10. Surendran, A. *et al.* Approaches to Generating Virtual Patient Cohorts with Applications in Oncology. in *Personalized Medicine Meets Artificial Intelligence* (eds. Cesario, A., D'Oria, M., Auffray, C. & Scambia, G.) 97–119 (Springer International Publishing, Cham, 2023). doi:10.1007/978-3-031-32614-1_8.

11. Craig, M., Gevertz, J. L., Kareva, I. & Wilkie, K. P. A practical guide for the generation of model-based virtual clinical trials. *Front. Syst. Biol.* **3**, 1174647 (2023).

12. Hormuth, D. A. *et al.* Mechanism-Based Modeling of Tumor Growth and Treatment Response Constrained by Multiparametric Imaging Data. *JCO Clinical Cancer Informatics* 1–10 (2019) doi:10.1200/CCI.18.00055.

13. Lazarou, G. *et al.* Integration of Omics Data Sources to Inform Mechanistic Modeling of Immune-Oncology Therapies: A Tutorial for Clinical Pharmacologists. *Clin Pharmacol Ther* **107**, 858–870 (2020).

14.     Arulraj, T. *et al.* Leveraging multi-omics data to empower quantitative systems pharmacology in immuno-oncology. *Briefings in Bioinformatics* **25**, bbae131 (2024).

15.     Stahlberg, E. A. *et al.* Exploring approaches for predictive cancer patient digital twins: Opportunities for collaboration and innovation. *Front. Digit. Health* **4**, 1007784 (2022).

16.     Cheng, Y. *et al.* Virtual Populations for Quantitative Systems Pharmacology Models. *Methods Mol Biol* **2486**, 129–179 (2022).

17.     Mellman, I., Chen, D. S., Powles, T. & Turley, S. J. The cancer-immunity cycle: Indication, genotype, and immunotype. *Immunity* **56**, 2188–2205 (2023).

18.     Chen, D. S. & Mellman, I. Oncology Meets Immunology: The Cancer-Immunity Cycle. *Immunity* **39**, 1–10 (2013).

19.     Niederer, S. A. *et al.* Creation and application of virtual patient cohorts of heart models. *Philos Trans A Math Phys Eng Sci* **378**, 20190558 (2020).

20.     Sové, R. J. *et al.* QSP-IO: A Quantitative Systems Pharmacology Toolbox for Mechanistic Multiscale Modeling for Immuno-Oncology Applications. *Clin. Pharmacol. Ther.* **9**, 484–497 (2020).

21.     Jafarnejad, M. *et al.* A Computational Model of Neoadjuvant PD-1 Inhibition in Non-Small Cell Lung Cancer. *AAPS J* **21**, 79 (2019).

22.     Ma, H. *et al.* A Quantitative Systems Pharmacology Model of T Cell Engager Applied to Solid Tumor. *AAPS J* **22**, 85 (2020).

23.     Ma, H. *et al.* Combination therapy with T cell engager and PD-L1 blockade enhances the antitumor potency of T cells as predicted by a QSP model. *J Immunother Cancer* **8**, e001141 (2020).

24.     Wang, H. *et al.* Conducting a Virtual Clinical Trial in HER2-Negative Breast Cancer Using a Quantitative Systems Pharmacology Model With an Epigenetic Modulator and Immune Checkpoint Inhibitors. *Front. Bioeng. Biotechnol.* **8**, 141 (2020).

25.     Wang, H., Ma, H., Sové, R. J., Emens, L. A. & Popel, A. S. Quantitative systems pharmacology model predictions for efficacy of atezolizumab and nab-paclitaxel in triple-negative breast cancer. *J Immunother Cancer* **9**, e002100 (2021).

26.     Anbari, S. *et al.* Using quantitative systems pharmacology modeling to optimize combination therapy of anti-PD-L1 checkpoint inhibitor and T cell engager. *Front. Pharmacol.* **14**, 1163432 (2023).

27.     Wang, H., Zhao, C., Santa-Maria, C. A., Emens, L. A. & Popel, A. S. Dynamics of tumor-associated macrophages in a quantitative systems pharmacology model of immunotherapy in triple-negative breast cancer. *iScience* **25**, 104702 (2022).

28.     Wang, H., Arulraj, T., Kimko, H. & Popel, A. S. Generating immunogenomic data-guided virtual patients using a QSP model to predict response of advanced NSCLC to PD-L1 inhibition. *npj Precis. Onc.* **7**, 55 (2023).

29.     Ippolito, A. *et al.* Eliciting the antitumor immune response with a conditionally activated PD-L1 targeting antibody analyzed with a quantitative systems pharmacology model. *CPT Pharmacom & Syst Pharma* psp4.13060 (2023) doi:10.1002/psp4.13060.

30.     Arulraj, T., Wang, H., Emens, L. A., Santa-Maria, C. A. & Popel, A. S. A transcriptome-informed QSP model of metastatic triple-negative breast cancer identifies predictive biomarkers for PD-1 inhibition. *Sci. Adv.* **9**, eadg0289 (2023).

31.     Gong, C., Ruiz-Martinez, A., Kimko, H. & Popel, A. S. A Spatial Quantitative Systems Pharmacology Platform spQSP-IO for Simulations of Tumor-Immune Interactions and Effects of Checkpoint Inhibitor Immunotherapy. *Cancers (Basel)* **13**, 3751 (2021).

32.     Ruiz-Martinez, A. *et al.* Simulations of tumor growth and response to immunotherapy by coupling a spatial agent-based model with a whole-patient quantitative systems pharmacology model. *PLoS Comput Biol* **18**, e1010254 (2022).

33.     Nikfar, M., Mi, H., Gong, C., Kimko, H. & Popel, A. S. Quantifying Intratumoral Heterogeneity and Immunoarchitecture Generated In-Silico by a Spatial Quantitative Systems Pharmacology Model. *Cancers* **15**, 2750 (2023).

34.     Zhang, S. *et al.* Integration of Clinical Trial Spatial Multi-omics Analysis and Virtual Clinical Trials Enables Immunotherapy Response Prediction and Biomarker Discovery. *Cancer Research* (2024) doi:10.1158/0008-5472.CAN-24-0943.

35.     Allen, R. J., Rieger, T. R. & Musante, C. J. Efficient Generation and Selection of Virtual Populations in Quantitative Systems Pharmacology Models. *CPT Pharmacometrics Syst Pharmacol* **5**, 140–6 (2016).

36.     Rieger, T. R. *et al.* Improving the generation and selection of virtual populations in quantitative systems pharmacology models. *Prog Biophys Mol Biol* **139**, 15–22 (2018).

37.     Mi, H. *et al.* Spatial and Compositional Biomarkers in Tumor Microenvironment Predicts Clinical Outcomes in Triple-Negative Breast Cancer. *bioRxiv* 2023.12.18.572234 (2023) doi:10.1101/2023.12.18.572234.

38.     Cimino-Mathews, A. *et al.* PD-L1 (B7-H1) expression and the immune tumor microenvironment in primary and metastatic breast carcinomas. *Hum Pathol* **47**, 52–63 (2016).

39.     Shiao, S. L. *et al.* Single-cell and spatial profiling identify three response trajectories to pembrolizumab and radiation therapy in triple negative breast cancer. *Cancer Cell* **42**, 70-84.e8 (2024).

40.     Jenner, A. L., Cassidy, T., Belaid, K., Bourgeois-Daigneault, M.-C. & Craig, M. In silico trials predict that combination strategies for enhancing vesicular stomatitis oncolytic virus are determined by tumor aggressivity. *J Immunother Cancer* **9**, e001387 (2021).

41.     Cardinal, O. *et al.* Establishing combination PAC-1 and TRAIL regimens for treating ovarian cancer based on patient-specific pharmacokinetic profiles using *in silico* clinical trials. *Comp Sys Onco* **2**, e1035 (2022).

42.     Limpert, E., Stahel, W. A. & Abbt, M. Log-normal Distributions across the Sciences: Keys and Clues. *BioScience* **51**, 341 (2001).

43.     Sender, R. *et al.* The total mass, number, and distribution of immune cells in the human body. *Proc. Natl. Acad. Sci. U.S.A.* **120**, e2308511120 (2023).

44.     Autissier, P., Soulas, C., Burdo, T. H. & Williams, K. C. Evaluation of a 12-color flow cytometry panel to study lymphocyte, monocyte, and dendritic cell subsets in humans. *Cytometry A* **77**, 410–9 (2010).

45.     Thorsson, V. *et al.* The Immune Landscape of Cancer. *Immunity* **48**, 812-830.e14 (2018).

46.     Garcia-Recio, S. *et al.* Multiomics in primary and metastatic breast tumors from the AURORA US network finds microenvironment and epigenetic drivers of metastasis. *Nat Cancer* **4**, 128–147 (2023).

47.     Rozenblatt-Rosen, O. *et al.* The Human Tumor Atlas Network: Charting Tumor Transitions across Space and Time at Single-Cell Resolution. *Cell* **181**, 236–249 (2020).

48.     Eddy, J. A. *et al.* CRI iAtlas: an interactive portal for immuno-oncology research. *F1000Res* **9**, 1028 (2020).

49.     Siegel, M. B. *et al.* Integrated RNA and DNA sequencing reveals early drivers of metastatic breast cancer. *Journal of Clinical Investigation* **128**, 1371–1383 (2018).

50.     Racle, J., De Jonge, K., Baumgaertner, P., Speiser, D. E. & Gfeller, D. Simultaneous enumeration of cancer and immune cell types from bulk tumor gene expression data. *eLife* **6**, e26476 (2017).

51.     Finotello, F. *et al.* Molecular and pharmacological modulators of the tumor immune contexture revealed by deconvolution of RNA-seq data. *Genome Med* **11**, 34 (2019).

52.     Venkatesh, K. P., Raza, M. M. & Kvedar, J. C. Health digital twins as tools for precision medicine: Considerations for computation, implementation, and regulation. *npj Digit. Med.* **5**, 150, s41746-022-00694–7 (2022).

53.      Laubenbacher, R., Mehrad, B., Shmulevich, I. & Trayanova, N. Digital twins in medicine. *Nat Comput Sci* **4**, 184–191 (2024).

54.      Katsoulakis, E. *et al.* Digital twins for health: a scoping review. *npj Digit. Med.* **7**, 77 (2024).

55.      Moingeon, P., Chenel, M., Rousseau, C., Voisin, E. & Guedj, M. Virtual patients, digital twins and causal disease models: Paving the ground for in silico clinical trials. *Drug Discovery Today* **28**, 103605 (2023).

56.      Vallée, A. Digital twin for healthcare systems. *Front. Digit. Health* **5**, 1253050 (2023).

57.      Wright, L. & Davidson, S. How to tell the difference between a model and a digital twin. *Adv. Model. and Simul. in Eng. Sci.* **7**, 13 (2020).

58.      An, G. & Cockrell, C. Drug Development Digital Twins for Drug Discovery, Testing and Repurposing: A Schema for Requirements and Development. *Front. Syst. Biol.* **2**, 928387 (2022).

59.      Laubenbacher, R. *et al.* Building digital twins of the human immune system: toward a roadmap. *npj Digit. Med.* **5**, 64 (2022).

60.      Wu, C. *et al.* MRI-Based Digital Models Forecast Patient-Specific Treatment Responses to Neoadjuvant Chemotherapy in Triple-Negative Breast Cancer. *Cancer Research* **82**, 3394–3404 (2022).

61.      Board on Mathematical Sciences and Analytics *et al. Opportunities and Challenges for Digital Twins in Biomedical Research: Proceedings of a Workshop-in Brief*. 26922 (National Academies Press, Washington, D.C., 2023). doi:10.17226/26922.

62.      Lorenzo, G. *et al.* Patient-Specific, Mechanistic Models of Tumor Growth Incorporating Artificial Intelligence and Big Data. *Annu Rev Biomed Eng* (2024) doi:10.1146/annurev-bioeng-081623-025834.

63.      Jarrett, A. M. *et al.* Quantitative magnetic resonance imaging and tumor forecasting of breast cancer patients in the community setting. *Nat Protoc* **16**, 5309–5338 (2021).

64.     Committee on Foundational Research Gaps and Future Directions for Digital Twins *et al.*

        *Foundational Research Gaps and Future Directions for Digital Twins*. 26894 (National Academies

        Press, Washington, D.C., 2024). doi:10.17226/26894.

65.     Alber, M. *et al.* Integrating machine learning and multiscale modeling—perspectives, challenges,

        and opportunities in the biological, biomedical, and behavioral sciences. *npj Digit. Med.* **2**, 115 (2019).

66.     Susilo, M. E. *et al.* Systems-based digital twins to help characterize clinical DOSE–RESPONSE and

        propose predictive biomarkers in a Phase I study of bispecific antibody, mosunetuzumab, in NHL.

        *Clinical Translational Sci* cts.13501 (2023) doi:10.1111/cts.13501.

67.     Tivay, A., Kramer, G. C. & Hahn, J.-O. Virtual Patient Generation using Physiological Models

        through a Compressed Latent Parameterization. in *2020 American Control Conference (ACC)* 1335–

        1340 (IEEE, Denver, CO, USA, 2020). doi:10.23919/ACC45564.2020.9147298.

68.     Sun, T., He, X. & Li, Z. Digital twin in healthcare: Recent updates and challenges. *DIGITAL HEALTH*

        **9**, 205520762211496 (2023).